\documentclass[graybox]{svmult}

\usepackage{mathptmx}       
\usepackage{helvet}         
\usepackage{courier}        
\usepackage{type1cm}        
\usepackage{makeidx}         
\usepackage{graphicx}        
\usepackage{multicol}        
\usepackage[bottom]{footmisc}

\makeindex             

\usepackage{fancyhdr}
\begin{document}

\title*{Does environment affect the star formation histories of early-type
galaxies?}
\titlerunning{Environment and elliptical galaxies}
\author{I. Ferreras, A. Pasquali, B. Rogers}
\authorrunning{Ferreras et al.} 
\institute{
  I. Ferreras \at MSSL, University College London, Holmbury St Mary,
  Dorking, Surrey RH5 6NT, UK. \email{ferreras@star.ucl.ac.uk}
  \and A. Pasquali \at ARI-Universit\"at Heidelberg,
  M\"onchhofstr. 12-14, D-69120 Heidelberg, Germany. \email{pasquali@ari.uni-heidelberg.de}
  \and B. Rogers \at Department of Physics, King's College London,
  Strand, London WC2R 2LS, UK.
}
%
%
\maketitle

\vskip-1.2truein

\abstract{Differences in the stellar populations of galaxies can be
  used to quantify the effect of environment on the star formation
  history. We target a sample of early-type galaxies from the Sloan
  Digital Sky Survey in two different environmental regimes: close
  pairs and a general sample where environment is measured by the mass
  of their host dark matter halo. We apply a blind source separation
  technique based on principal component analysis, from which we
  define two parameters that correlate, respectively, with the average
  stellar age ($\eta$) and with the presence of recent star formation
  ($\zeta$) from the spectral energy distribution of the galaxy. We
  find that environment leaves a second order imprint on the spectra,
  whereas local properties -- such as internal velocity dispersion --
  obey a much stronger correlation with the stellar age
  distribution.}

\section{The environment of early-type galaxies}
\label{sec:intro}

Within the standard framework of structure formation, galaxies grow in
a hierarchical fashion from small structures, progressively merging
into more massive systems. Galaxies in regions with a higher
over-density will collapse earlier than galaxies in under-dense
regions. Hence, we expect a significant dependence of the star
formation histories of galaxies with the environment were they form
and evolve. The morphology-density relation \cite{ad80} whose 30 year
anniversary we celebrate in this Symposium is indeed proof of the fact
that environmental mechanisms are important in shaping the galaxy
populations we see today. This contribution focuses on the star formation
histories of a type of galaxies that are especially sensitive tracers
of environment.

The dynamical state of elliptical galaxies suggests a formation
process driven by galaxy-galaxy interactions. The current
interpretation for their formation history involves major mergers,
although an observational quantification of the role of mergers and
their impact on the underlying stellar populations remains an open
question \cite{cons,eGDS}. This morphological type is thus an optimal
target to understand the effect of environment on galaxy formation.
In order to quantify the effect of environment, we compare large data
sets comprising spectral energy distributions of early-type galaxies
in different environmental regimes. Differences in their properties
are determined by the application of a blind-source separation method
whereby the data alone -- no modelling -- are used to define an
observable that discriminates between galaxies based on their
spectral information.

The work presented here is based on spectra from the Sloan Digital Sky
Survey (SDSS, \cite{sdss}). By applying Principal Component Analysis (PCA) to the
spectra, we find that most of the information locked in the data (over
99\% in the sense of variance) resides in the first two principal
components. Projecting all spectra on to a rotated version of these
two components allowed us to generate two PCA-based parameters: $\eta$
and $\zeta$. The rotation of components in the parameter space spanned
by the eigenvectors of the covariance matrix is often used as a method
to go beyond a simple decorrelation of the data, such as independent
component analysis, where statistical independence between components
is sought (see e.g. \cite{ica}). By comparing the rotated components
with models of population synthesis for a wide range of star formation
histories, we find that the projection of a galaxy spectrum on to the
$\eta$ component correlates with average age for metal-rich
populations (typical of the types of galaxies explored here).
Furthermore, projections on to the $\zeta$ component are very
sensitive to the presence of recent star formation, as found when
matching against NUV fluxes of early-type galaxies from GALEX (see
\cite{hcg,pcaEs} for details).

Two environmental regimes are studied. In the next section we consider
close pairs of interacting early-type galaxies. In \S3 we present the
results for a general, volume-limited sample of early-type galaxies
out to z$<$0.1, where environment is parameterised according to the mass
of the halo where the galaxy resides, using the groups catalogue from
\cite{gcat}.

\begin{figure}[t]
\begin{center}
\includegraphics[scale=0.45]{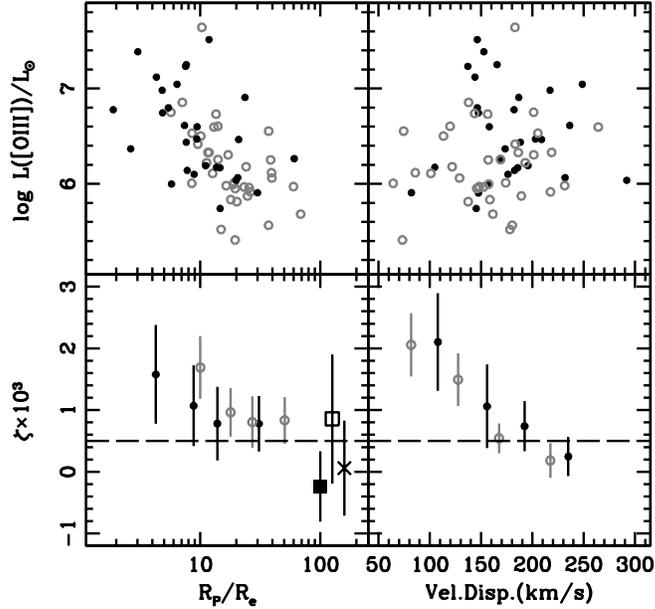}
\caption{Spectral properties of the close pairs sample, given by the
  [OIII] luminosity ({\sl top}) and PCA parameter $\zeta$, that tracks the
  underlying stellar populations ({\sl bottom}). Values of $\zeta>500$
  (horizontal dashed line) reflect a significant amount of star
  formation. The black solid and grey open dots represent galaxies with (without)
  visual signs of interaction (as seen on the SDSS-DR6 images). The
  dots in the top panel represent individual galaxies (where AGN-like
  emission is found), whereas the bottom panels give the average and
  error on the average for subsamples binned according to pair
  separation (R$_{\rm P}$, {\sl left}), given in units of the effective
  radius) or velocity dispersion ({\sl right}). The solid and hollow squares
  in the bottom-left panel shows the average and RMS of a field
  distribution of NUV faint and bright early-type galaxies,
  respectively, and the cross shows the average value, regardless of
  NUV flux.}
\label{fig:pairs} 
\end{center}
\end{figure}

\section{Environment over small scales: Close pairs}
\label{sec:pairs}

We define a sample of close pairs by choosing galaxies from SDSS with
a radial velocity difference below 500 km/s and a physical separation
within 30 kpc along the transversal direction. This sample is then
visually classified, selecting only those systems consisting of both
members being early-type galaxies. Our final sample comprises 347
pairs, and is described in \cite{pairs}. This definition of a pair
produces a clean sample to probe the details of the triggering of star
formation and the onset of nuclear emission, as the progenitors are
expected to have neither star formation nor AGN activity. The top
panels of figure~\ref{fig:pairs} shows the luminosity of the [OIII]
line for galaxies with AGN-like emission (which amounts to 9.5\% of
the total sample). A trend is found with respect to pair separation
({\sl left}, given in units of the effective radius of each galaxy),
but no correlation is found with internal velocity dispersion ({\sl
  right}). The bottom panels show the distribution of the $\zeta$
PCA-component, which tracks recent star formation. In this case, the
general sample (not only AGN) is shown, with the dots and error bars
representing the average and its error on subsamples binned according
to separation or velocity dispersion. The decomposition into PCA
components is explained in detail in \cite{pcaEs}. As a comparison,
the squares in the bottom left panel give the values of $\zeta$ for a
field sample of early-types separated with respect to their NUV flux
-- a direct tracer of recent star formation.  One can apply a
threshold at $\zeta > 500$ above which most galaxies will have
undergone recent star formation (horizontal dashed line).  The
comparison shows that early-type galaxies in close pairs have a higher
rate of recent star formation than the general sample. The trend with
respect to pair separation is visible, although weaker than the
correlation found with respect to a more intrinsic observable, namely
velocity dispersion. The observed onset of star formation in close
pairs, even before the galaxies display any visual feature of
interaction, can be explained by the presence of clouds of gas within
their halos, as observed by, e.g.\cite{gasE}. Even in galaxies
separated over 100 effective radii, we find that their level of recent
star formation is higher than the average value found for a general
sample of (non close pair) early-type galaxies, shown by the '$\times$' sign
on the bottom-left panel of figure~\ref{fig:pairs}. Furthermore, the
strong correlation between pair separation and AGN activity -- traced
by the luminosity of the [OIII] line -- suggests that the triggering
of star formation precedes AGN activity (see \cite{pairs} for details)

\begin{figure}[t]
\begin{center}
\includegraphics[scale=0.45]{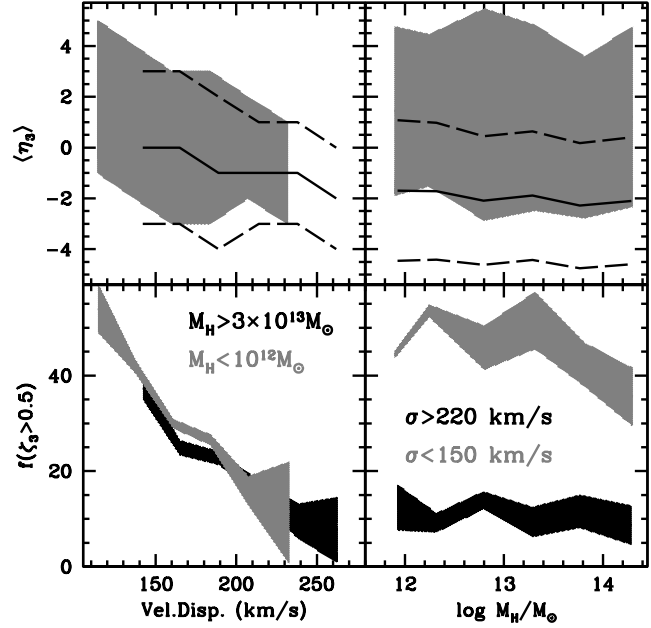}
\caption{Effect of group environment on the stellar populations
of early-type galaxies. The top panels show the value of PCA
parameter $\eta_3\equiv\eta\times 10^3$, which maps average stellar age.
The bottom panels give the fraction of galaxies with recent star
formation, as characterized by the PCA parameter $\zeta_3\equiv\zeta\times 10^3>0.5$. 
The samples are shown with respect to a local property (velocity
dispersion, {\sl left}) and an environment-related property (group halo
mass, {\sl right}).  }
\label{fig:groups} 
\end{center}
\end{figure}

\section{Environment over large scales: Groups}
\label{sec:groups}

For a more general description of environment, we use a sample of SDSS
early-type galaxies from \cite{Ber} and define their environment
according to the host dark matter halo, given in the groups
catalogue of \cite{gcat}. This definition of ``group'' is more general
than the traditional concept of galaxy groups, and can extend from
isolated galaxies -- where one dark matter halo contains only one
galaxy, to clusters. However, our sample treats galaxies within 
a group in the same way, so that we do not resolve, for instance, the
environmental differences between the central region and the outskirts
of a galaxy cluster. Rather, we want to quantify the importance of the
background density where the galaxy lives, on its past and recent
star formation history.

Each galaxy is described by a ``local'' observable (their
central velocity dispersion) and by an ``environmental'' parameter
(the host halo mass). Figure~\ref{fig:groups} shows the trend of the
PCA-based parameters ($\eta$ and $\zeta$, with the subindex '3' meaning
$\times 10^3$), with respect to the local and environmental
observables. The shaded areas map the RMS of the distribution,
segregated with respect to halo mass or velocity dispersion, as
labelled. For clarity, the black shaded area in the top panels is
replaced by solid and dashed black lines, tracing the mean and RMS,
respectively. The bottom panel shows the fraction of galaxies with a
value of $\zeta$ above the 500 threshold for which the galaxy is
assumed to have undergone recent star formation (see the comparison in
the bottom left panel of figure~\ref{fig:pairs} between $\zeta$ and
NUV bright/faint galaxies). The figure suggests local properties, such
as velocity dispersion, play an important role in shaping the
underlying stellar populations, both for the average age ($\eta$, top
panels) and for the amount of recent star formation ($\zeta$, bottom
panels). Environment, as defined here by halo mass, only gives a
correction to this trend, with galaxies in less massive halos
appearing with a similar average age, but with an slightly higher amount
of recent star formation at fixed velocity dispersion. We should
emphasize here that this definition of environment is adequate for
large scale structure analyses. The more acute effects of environment,
such as those encountered when galaxies fall into the potential wells
of clusters, are not accounted for in this definition.

%
%


\end{document}